\documentclass[12pt,floatfix]{revtex4}
\usepackage{graphicx}
\usepackage{amsmath}

\begin{document}

\title{Conformational properties of neutral and charged alanine and glycine chains}

\author{Alexander V. Yakubovitch\dag,
        Ilia A. Solov'yov\dag\ddag, 
        Andrey V. Solov'yov\dag\ddag, 
        and Walter Greiner\ddag}

\address{\dag\ A.F. Ioffe Physical-Technical Institute, Russian
Academy of Sciences, Politechnicheskaja str. 26, 194021 St Petersburg,  Russia}  

\address{\ddag\ Institut f\"{u}r Theoretische Physik der Universit\"{a}t
Frankfurt am Main, Robert-Mayer Str. 8-10, D-60054 Frankfurt am Main, Germany}

\email{solovyov@th.physik.uni-frankfurt.de,ilia@th.physik.uni-frankfurt.de}

\begin{abstract}
We have investigated the potential energy surface for neutral and charged alanine
and glycine chains consisting of up to 6 amino acids.
For these molecules we have calculated potential energy surfaces as a function
of the Ramachandran angles $\varphi$ and $\psi$.
Our calculations are performed within {\it ab initio} theoretical framework based on
the density functional
theory and also within semi-empirical model approaches. We have demonstrated that the excessive
positive charge of the system influences strongly its geometrical and conformational properties.
With increasing of the excessive charge  amino acid chains become unstable and decay into two or
more fragments. We have analysed how the secondary structure of polypeptide chains influences
the formation of the potential energy landscapes. We have calculated the energy barriers for
transitions between different molecular conformations and determined the ones being
energetically the most favourable. 

\end{abstract}

\maketitle

\section{Introduction}

{\it This work was presented on "The eighth European Conference on Atomic and Molecular Physics"
(ECAMPVIII) (Rennes, France, July 6-10, 2004) and on the 
"Electronic Structure Simulations of Nanostructures" workshop (ESSN2004)
(Jyv\"askyl\"a Finland, June 18-21, 2004).}
\\
\\

Amino acids are building blocks for proteins. Recently, it became possible to study experimentally
fragments of proteins, i.e. chains of amino acids, in a gas phase with the use of the MALDI
mass spectroscopy \cite{Beavis96,Cohen96,Karas88}.
{\it Ab initio} theoretical investigations of amino acid chains began also only recently
\cite{Salahub01,Pliego03,Srinivasan02,Head-Gordon91,Gould94,Beachy97,LesHouches}
and are still in its infancy.

We have investigated the potential energy surface for neutral and charged alanine
and glycine chains consisting of up to 6 amino acids. For these molecules we have
calculated potential energy surfaces as a function of the Ramachandran angles
$\varphi$ and $\psi$ often used for the characterization of the polypeptide chains
\cite{Gunasekaran96,Protbase}. Our calculations are performed within {\it ab initio}
theoretical framework based on the density functional theory and also within semi-empirical
model approaches. We have demonstrated that the excessive positive charge of the system
influences strongly its geometrical and conformational properties. With increasing of the
excessive charge  amino acid chains become unstable and decay into two or more fragments.
We have analysed how the secondary structure of polypeptide chains influences the formation
of the potential energy landscapes. We have calculated the energy barriers for transitions
between different molecular conformations and determined the ones being energetically
the most favourable.

\section{Theoretical methods}

Our exploration of the potential energy surface of alanine and glycine chains
is based on the density-functional theory (DFT) accounting for all electrons in
the system and on the semiempirical AM1 method \cite{Dewar77,Davis81,Anders93}.

Within the DFT one has to solve the Kohn-Sham equations, which read as
(see e.g. \cite{MetCl99,LesHouches}):

\begin{equation}
\left( \frac{\hat p^2}2+U_{ions}+V_{H}+V_{xc}\right)
\psi_i =\varepsilon _i \psi _i,
\end{equation}
where the first term represents the kinetic energy of the $i$-th electron,
and $U_{ions}$ describes its attraction to the ions in the cluster,
$V_{H}$ is the Hartree part of the interelectronic interaction:
\begin{equation}
V_{H}(\vec r)=\left. \int \frac{\rho(\vec r\,')}{|\vec r-\vec r\,'|}
\, d\vec r\,'\right.,
\end{equation}
and $\rho(\vec r\,')$ is the  electron density:
\begin{equation}
\rho(\vec r)=\sum_{\nu=1}^{N} \left|\psi_i(\vec r) \right|^2,
\end{equation}

\noindent
where $V_{xc}$ is the local exchange-correlation potential,
$\psi_i$ are the electronic orbitals and $N$ is the number of 
electrons in the cluster.

The exchange-correlation potential is defined as the functional
derivative of the exchange-correlation energy functional:
\begin{equation}
V_{xc}=\frac{\delta E_{xc}[\rho]}{\delta \rho(\vec r)},
\end{equation}

The approximate functionals employed by DFT methods partition the
exchange-correlation energy into two parts, referred to as exchange
and correlation parts. Both parts are the functionals of
the electron density, which can be of two distinct types: either local functional
depending on only the electron density $\rho$ or gradient-corrected functionals depending on both
$\rho$ and its gradient, $\nabla\rho$. In literature, there is a variety of exchange correlation
functionals. In our work we use the Becke's three parameter gradient-corrected exchange
functional with the gradient-corrected correlation functional of Lee, Yang and Parr (B3LYP)
\cite{Becke88,LYP,Parr-book}. 
We utilize the standard 6-311++G(d,p) and 6-31G(2d,p)
basis set to expand the electronic orbitals $\psi_i$.

\section{Results of calculation}
\begin{figure}[h]
\includegraphics[scale=0.71,clip]{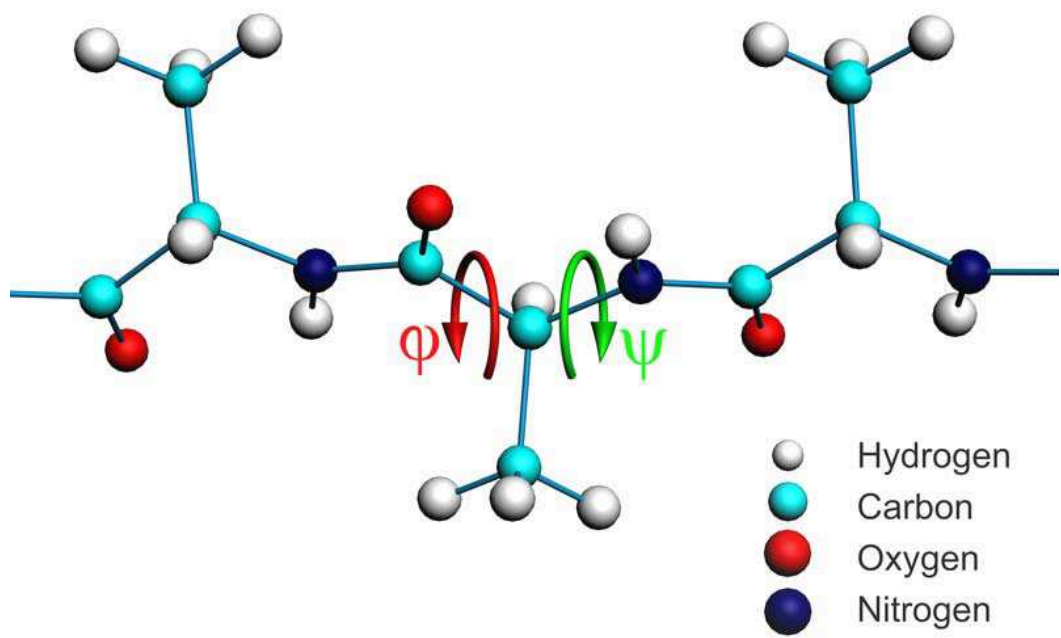}
\caption{Dihedral angles $\varphi$ and $\psi$ that are used to characterize the potential surface
of the polypeptide chain}
\label{angles}
\end{figure}

In figure \ref{angles} we show the dihedral angles $\varphi$ and $\psi$ that are used to
characterize the potential surface of the polypeptide chain.

\begin{figure}[h]
\includegraphics[scale=0.71,clip]{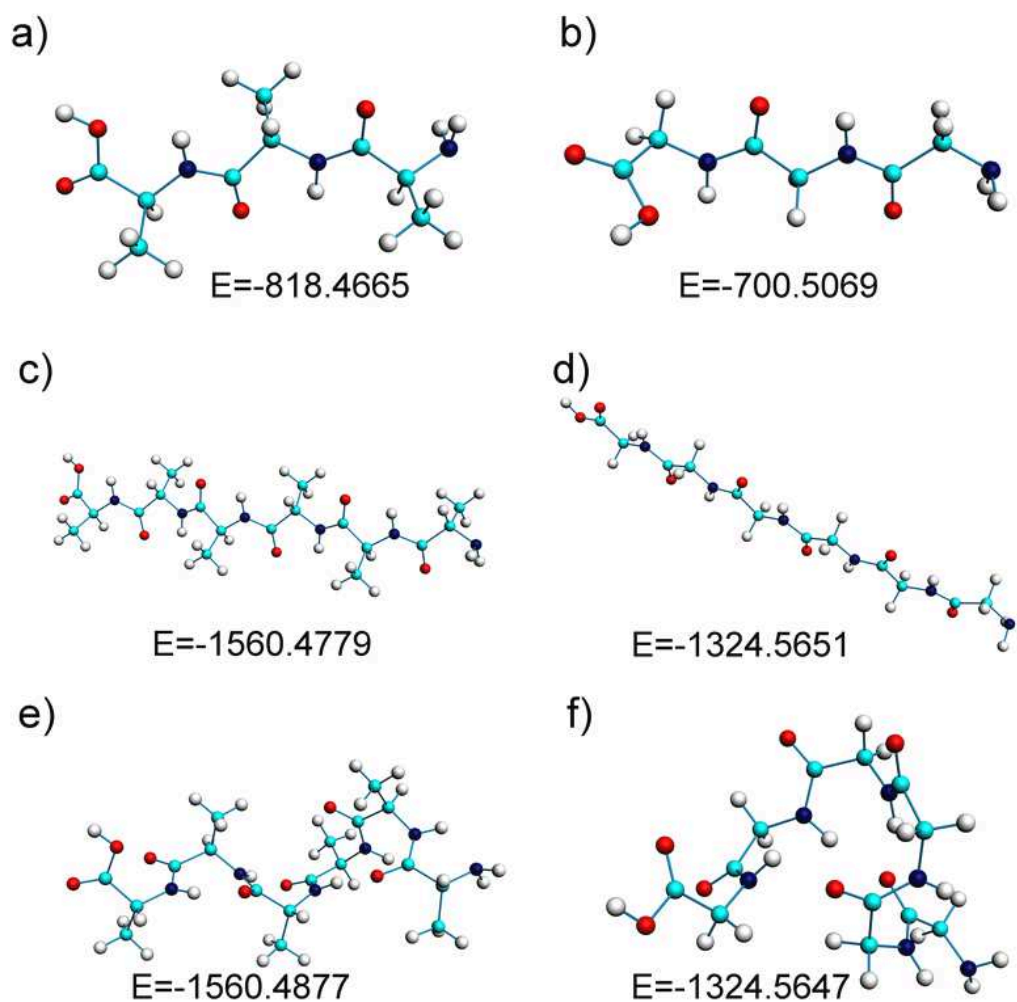}
\caption{Optimized geometries of the alanine and glycine polypeptide chains
calculated with the B3LYP/6-31++G(d,p) (a and b) and B3LYP/6-31G(2d,p) (c-f).
a) Alanine tripeptide;
b) Glycine tripeptide;
c) Alanine hexapeptide in sheet conformation;
d) Glycine hexapeptide in sheet conformation;
e) Alanine hexapeptide in helix conformation;
f) Glycine hexapeptide in helix conformation;
The energies below each image are given in a.u.}
\label{geom_neutral}
\end{figure}

In figure \ref{geom_neutral} we present the optimized geometries of the alanine and glycine
polypeptide chains that have been used for the exploration of the potential energy surface.
All geometries were optimized with the B3LYP density functional.
We used the 6-31++G(d,p) and 6-31G(2d,p) basis sets to expand the electronic orbitals in the
molecule.

\begin{figure}[h]
\includegraphics[scale=0.8,clip]{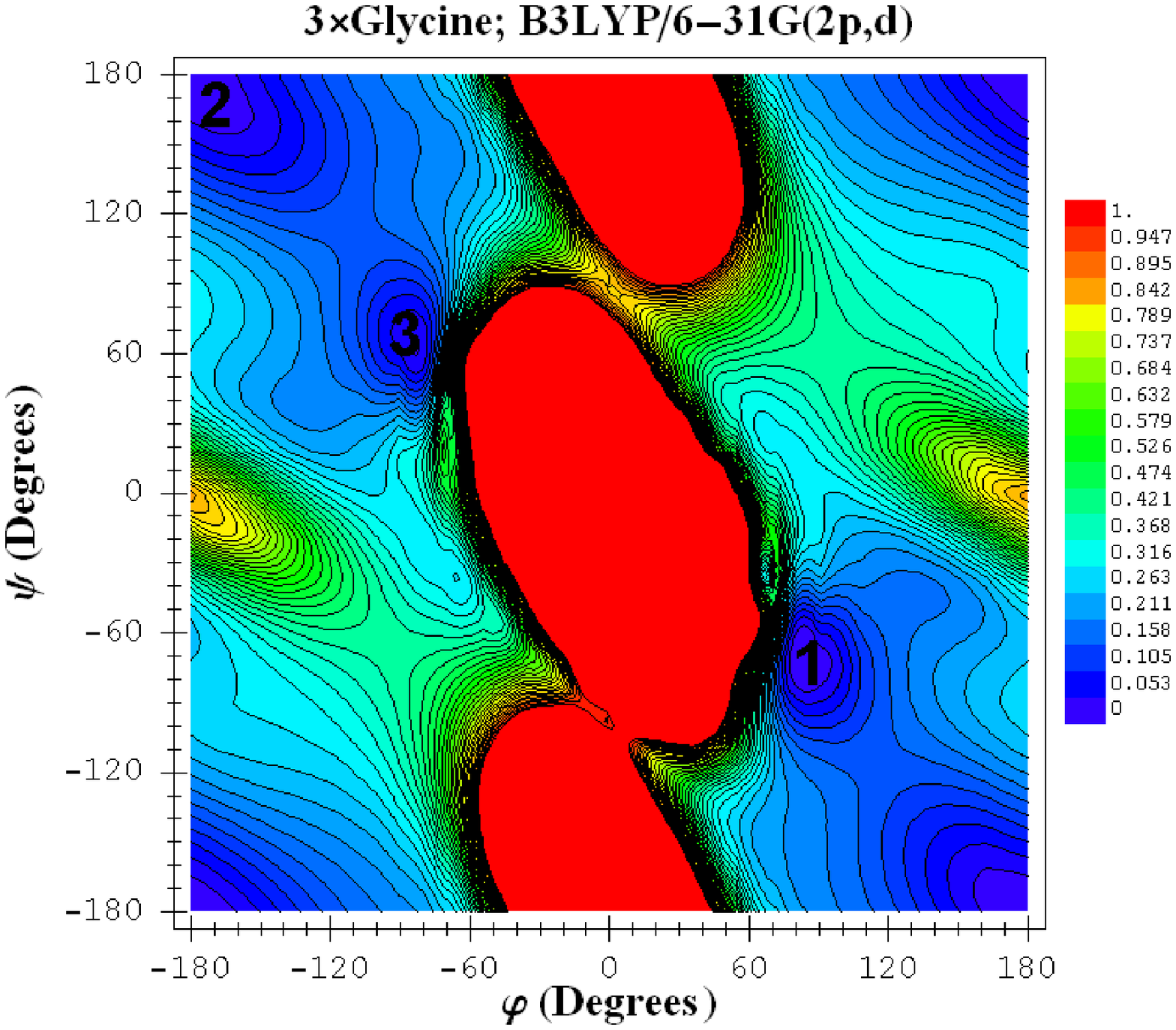}
\caption{Potential energy surface for the glycine tripeptide
calculated with the B3LYP/6-31G(2d,p) method.
Energies are given in eV. Numbers
mark energy minima on the potential energy surface.}
\label{map_gly_x3}
\end{figure}

\begin{figure}[h]
\includegraphics[scale=0.8,clip]{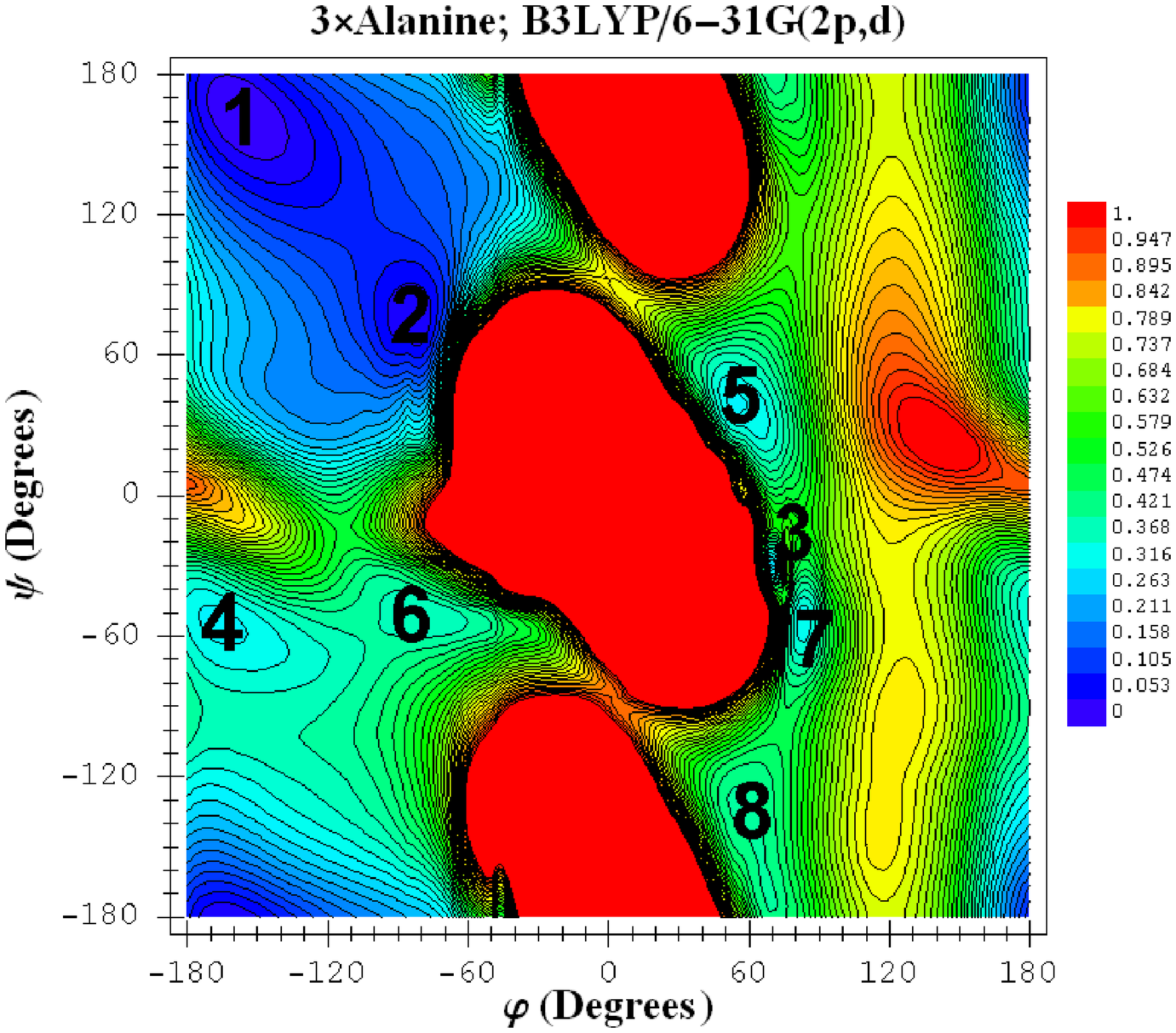}
\caption{Potential energy surface for the alanine tripeptide
calculated with the B3LYP/6-31G(2d,p) method. Energies are given in eV. Numbers
mark energy minima on the potential energy surface.}
\label{map_ala_x3}
\end{figure}

\begin{figure}[h]
\includegraphics[scale=0.8,clip]{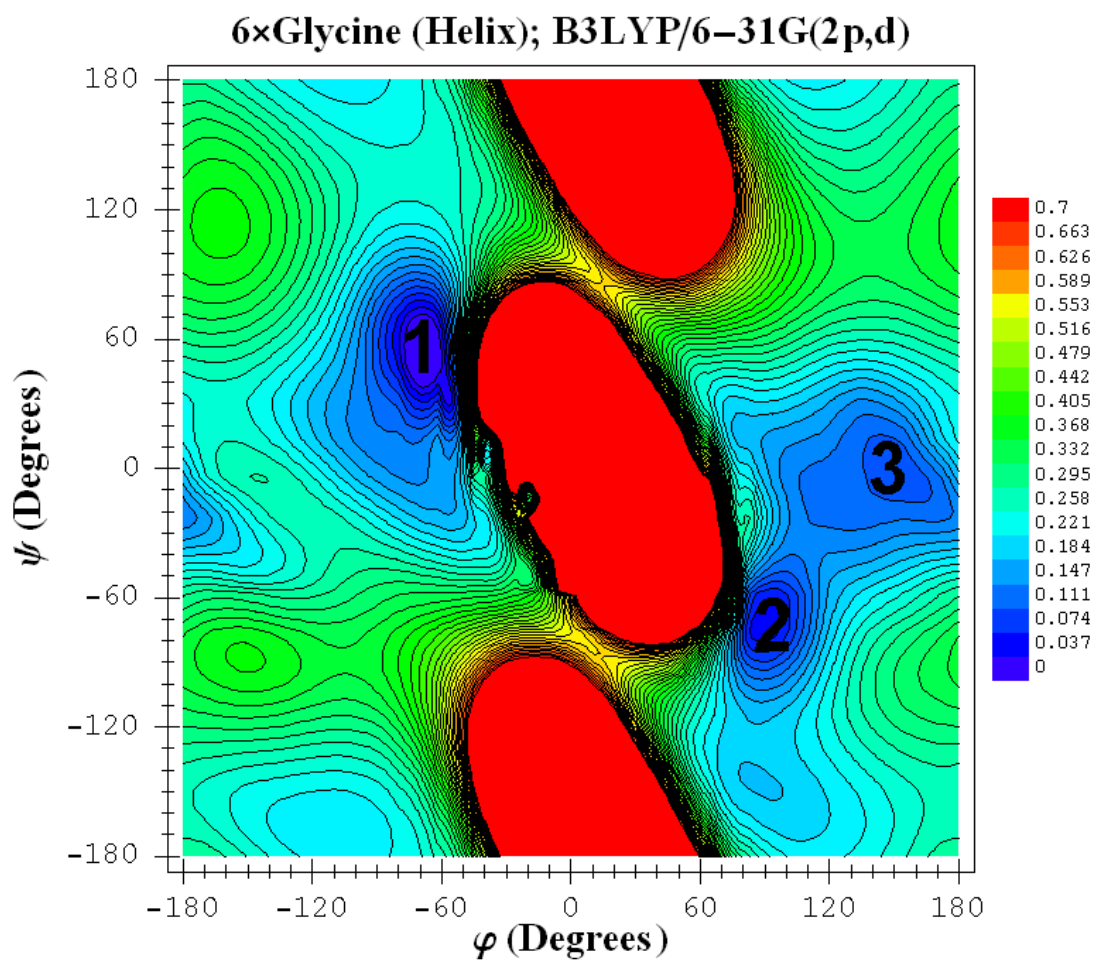}
\caption{Potential energy surface for the glycine hexapeptide
in helix conformation calculated with the B3LYP/6-31G(2d,p) method.
Energies are given in eV. Numbers
mark energy minima on the potential energy surface.}
\label{map_gly_x6_helix}
\end{figure}

\begin{figure}[h]
\includegraphics[scale=0.8,clip]{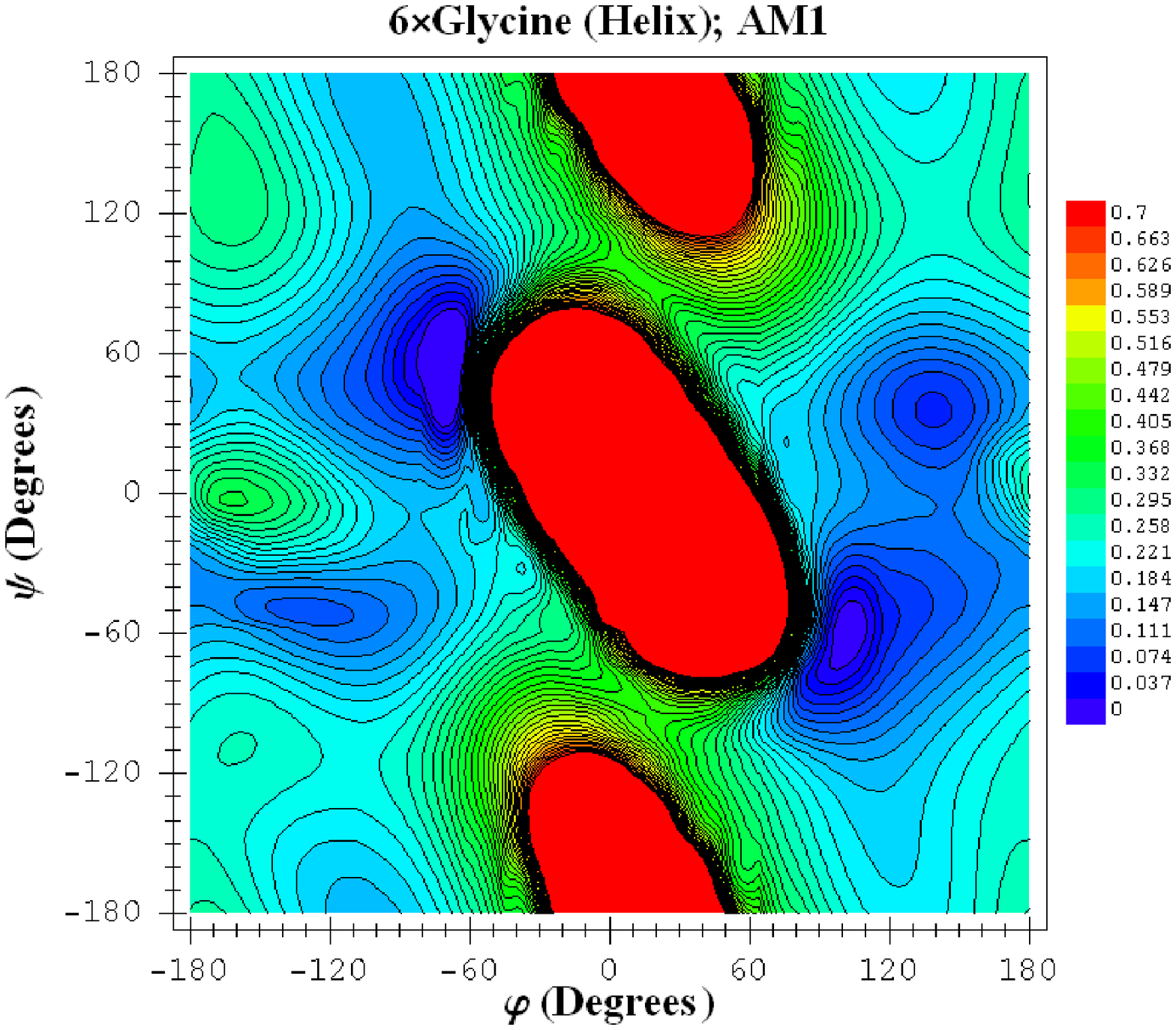}
\caption{Potential energy surface for the glycine hexapeptide
in helix conformation calculated with the AM1 method.
Energies are given in eV.}
\label{map_gly_x6_helix_am1}
\end{figure}

\begin{figure}[h]
\includegraphics[scale=0.8,clip]{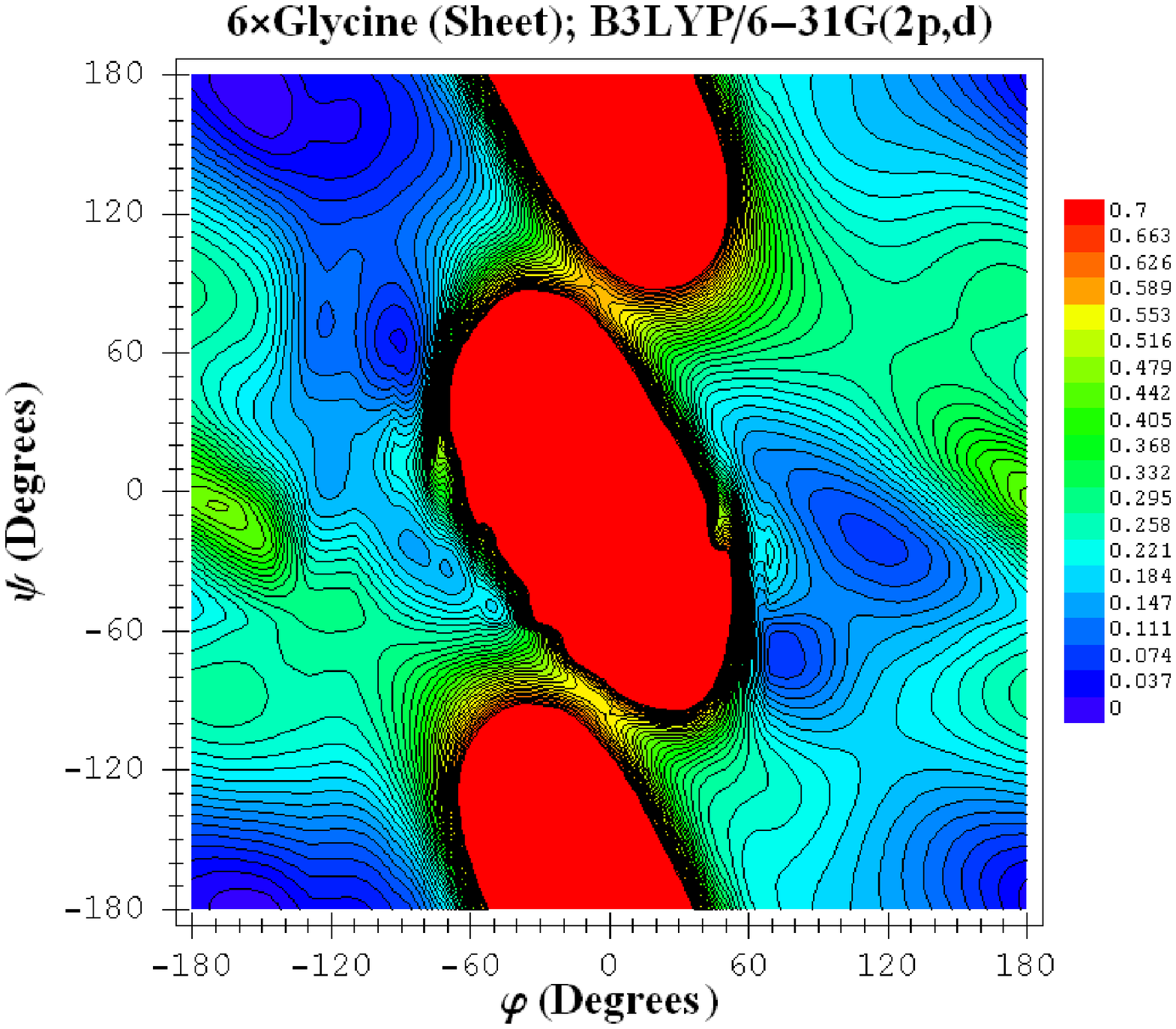}
\caption{Potential energy surface for the glycine hexapeptide
in sheet conformation calculated with the B3LYP/6-31G(2d,p) method.
Energies are given in eV.}
\label{map_gly_x6_sheet}
\end{figure}

\begin{figure}[h]
\includegraphics[scale=0.8,clip]{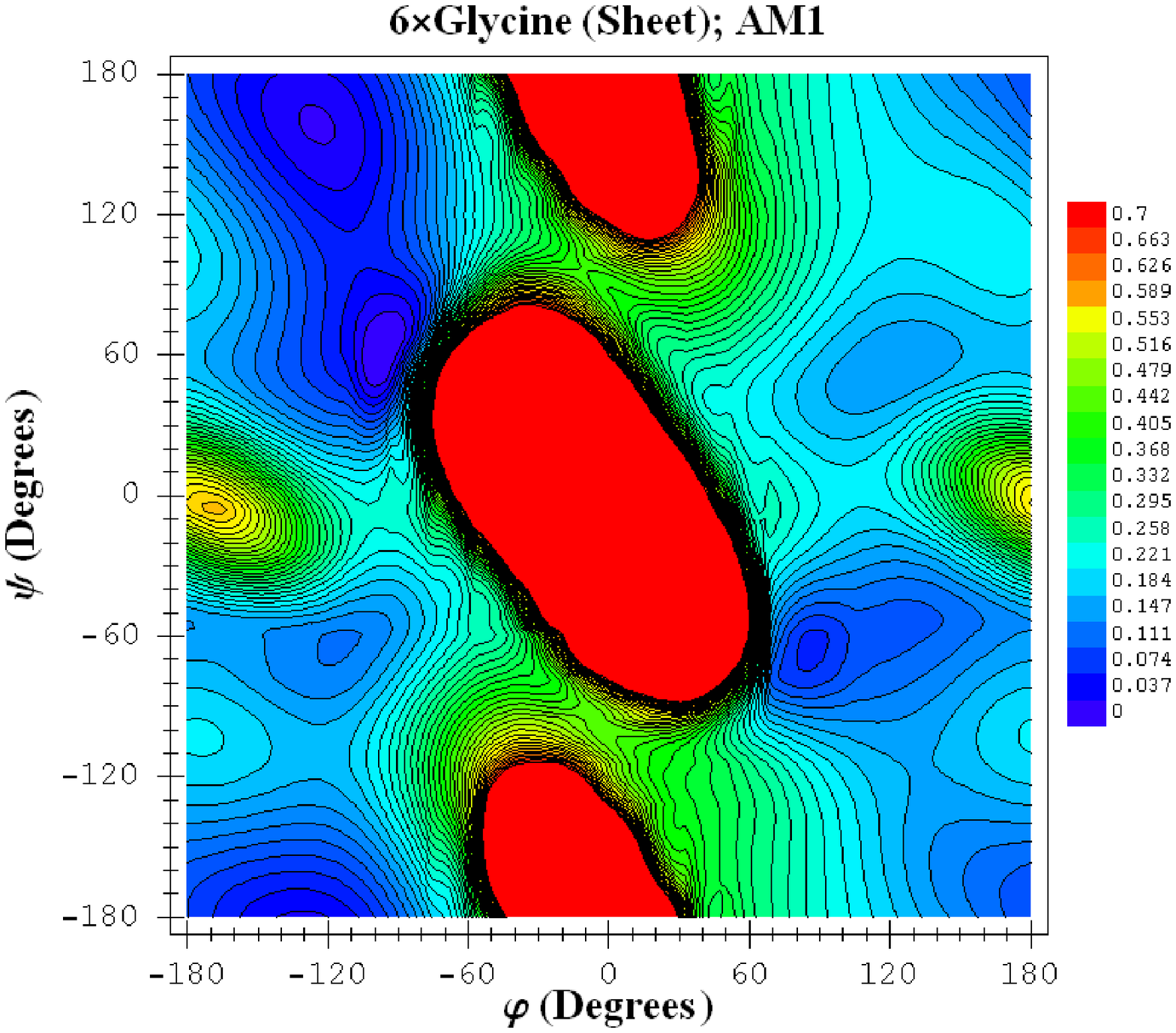}
\caption{Potential energy surface for the glycine hexapeptide
in sheet conformation calculated with the AM1 method.
Energies are given in eV.}
\label{map_gly_x6_sheet_am1}
\end{figure}

\begin{figure}[h]
\includegraphics[scale=0.8,clip]{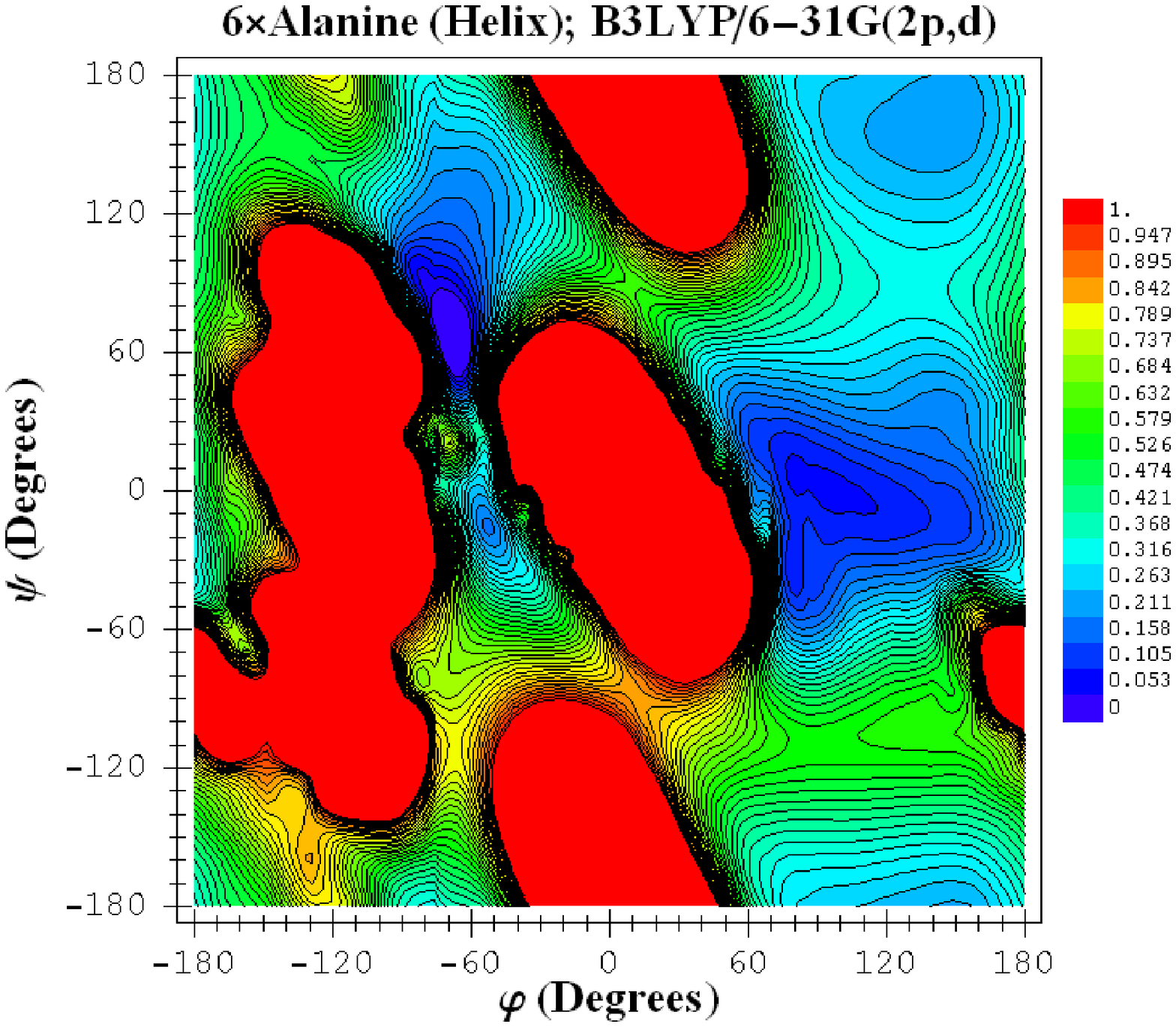}
\caption{Potential energy surface for the alanine hexapeptide
in helix conformation calculated with the B3LYP/6-31G(2d,p) method.
Energies are given in eV.}
\label{map_ala_x6_helix(right)}
\end{figure}

\begin{figure}[h]
\includegraphics[scale=0.8,clip]{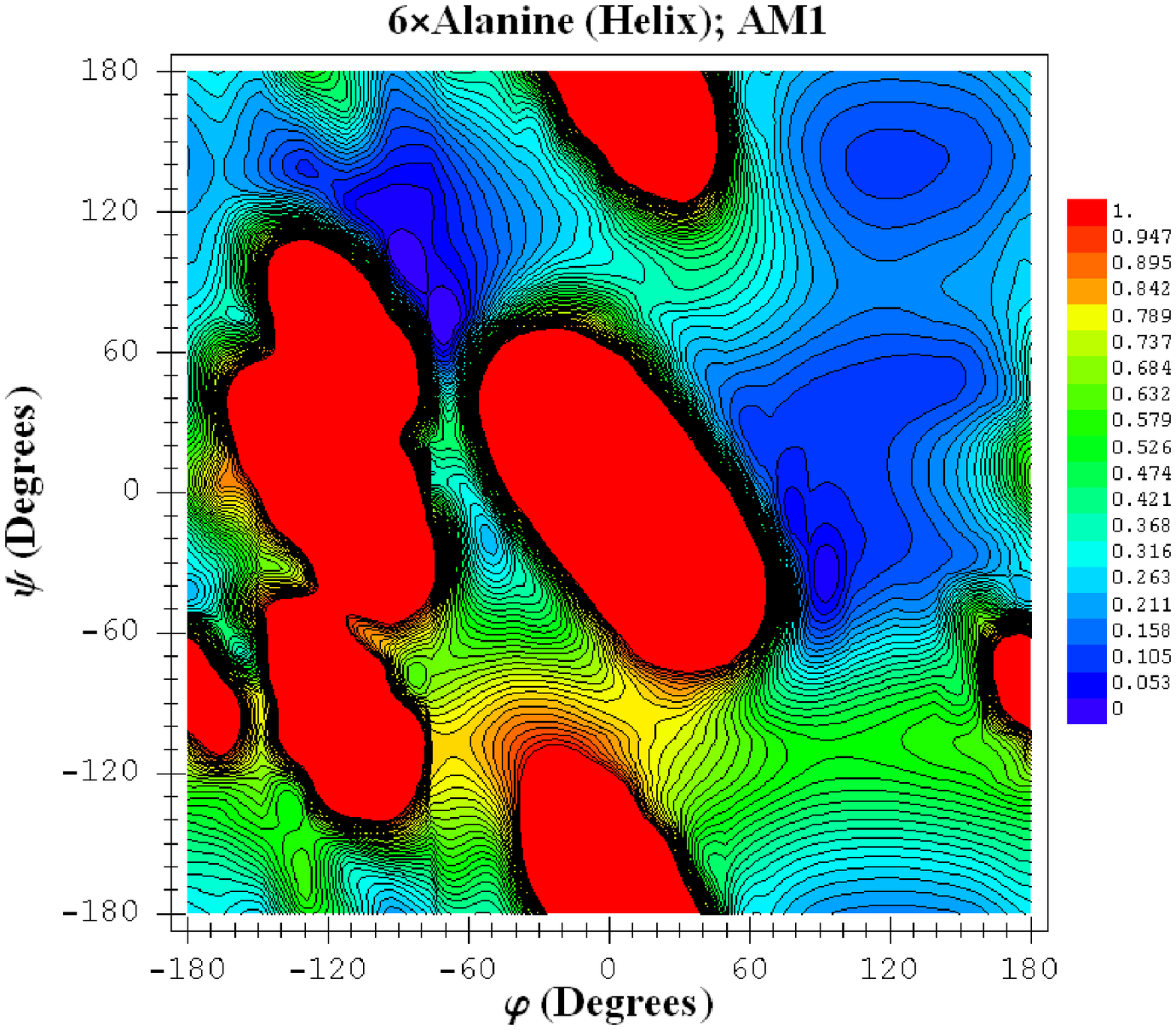}
\caption{Potential energy surface for the alanine hexapeptide
in helix conformation calculated with the AM1 method.
Energies are given in eV.}
\label{map_ala_x6_helix(right)_am1}
\end{figure}

\begin{figure}[h]
\includegraphics[scale=0.8,clip]{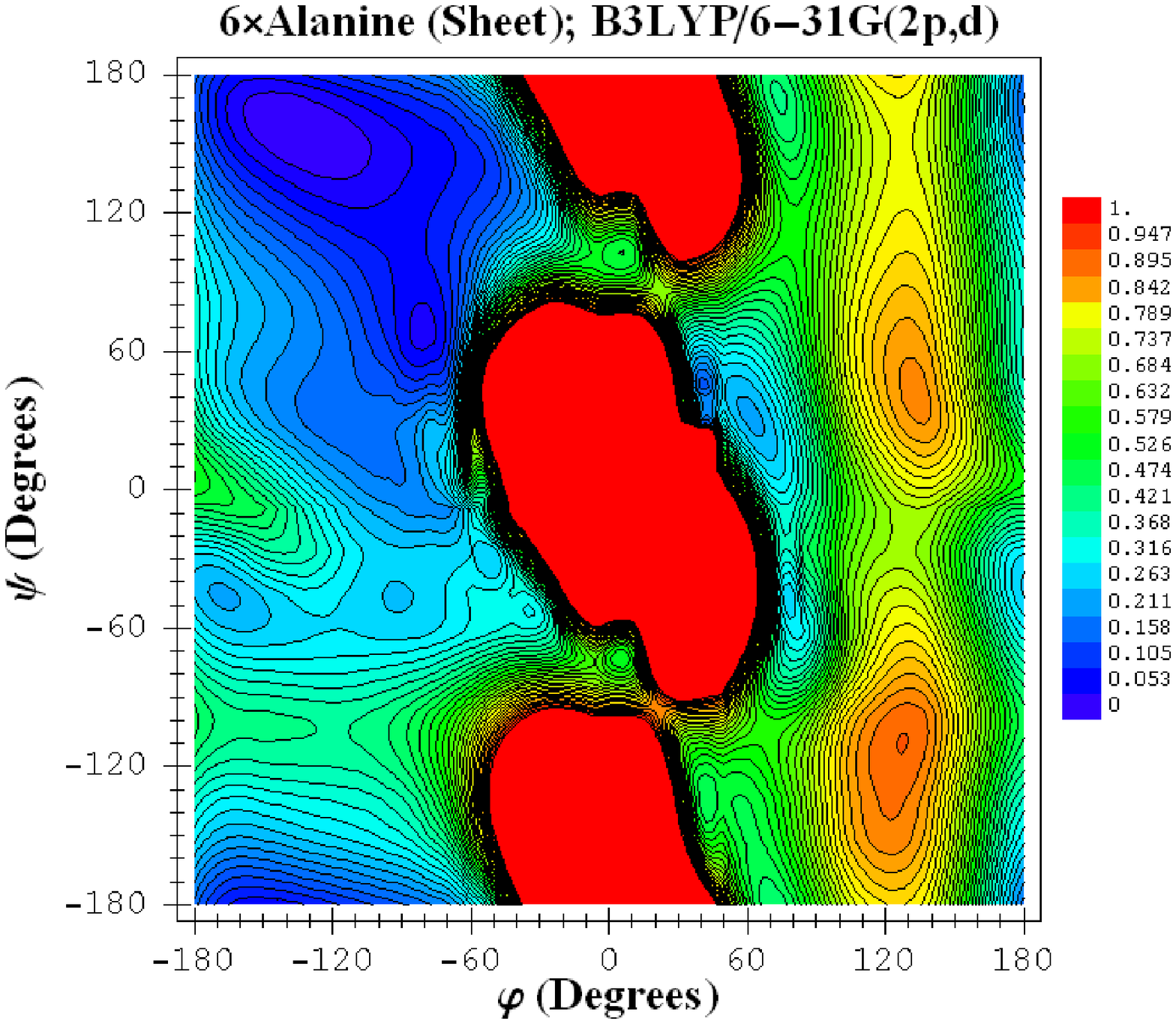}
\caption{Potential energy surface for the alanine hexapeptide
in sheet conformation calculated with the B3LYP/6-31G(2d,p) method.
Energies are given in eV.}
\label{map_ala_x6_sheet}
\end{figure}

\begin{figure}[h]
\includegraphics[scale=0.8,clip]{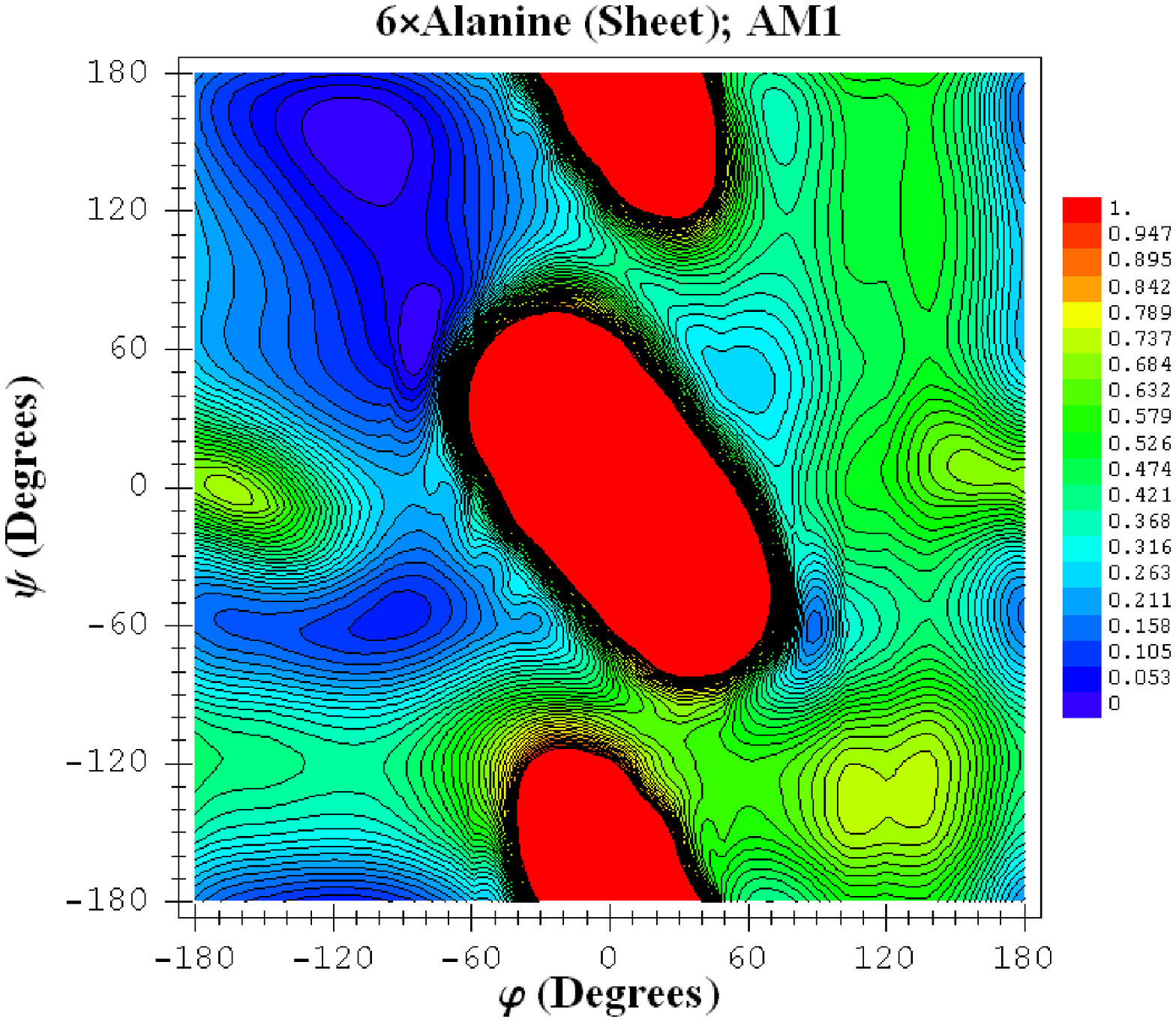}
\caption{Potential energy surface for the alanine hexapeptide
in sheet conformation calculated with the AM1 method.
Energies are given in eV.}
\label{map_ala_x6_sheet_am1}
\end{figure}

In figures \ref{map_gly_x3}-\ref{map_ala_x6_sheet_am1} we present the potential energy surfaces
for the polypeptide chains presented in figure \ref{geom_neutral}.

\begin{figure}[h]
\includegraphics[scale=0.7,clip]{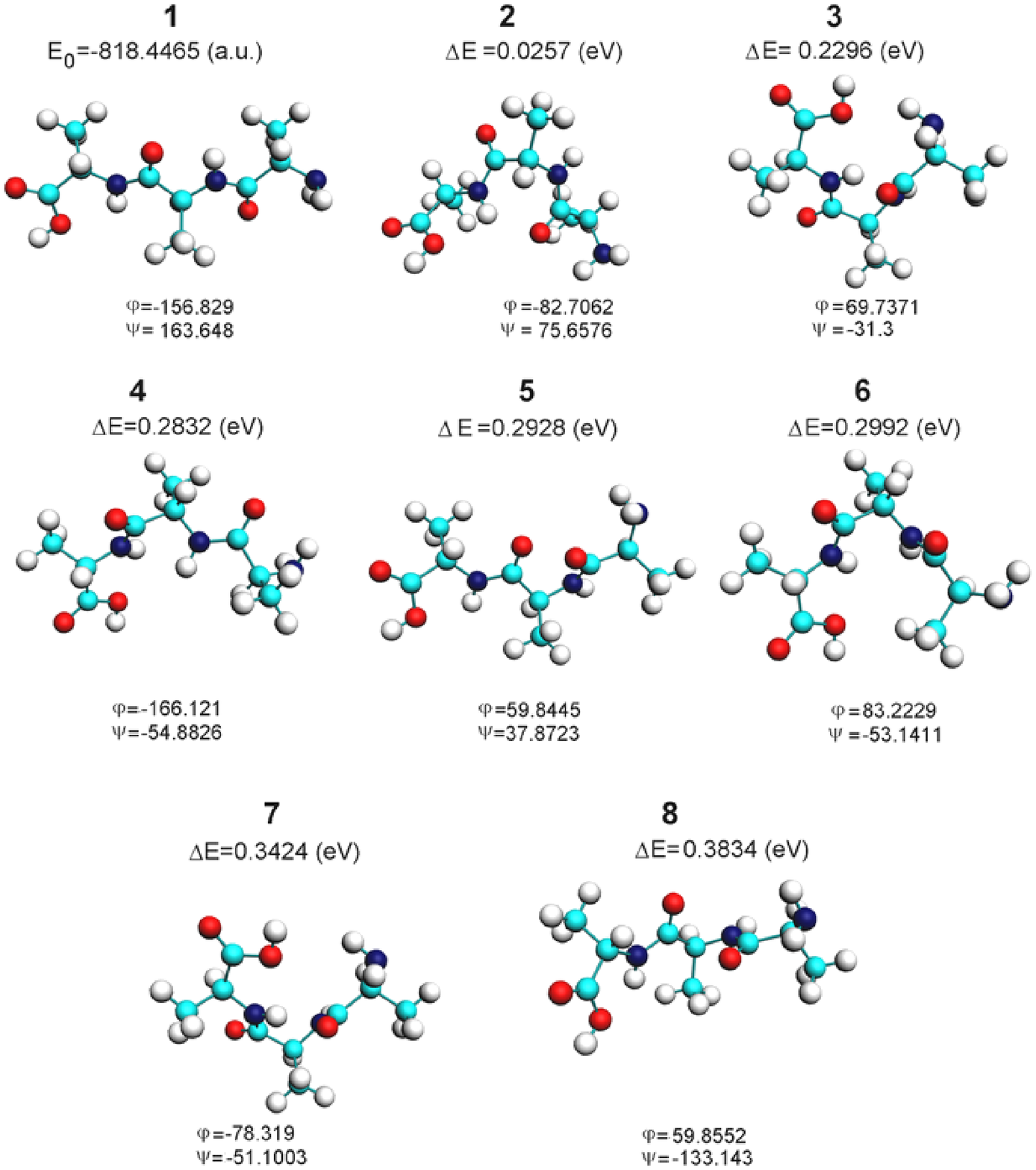}
\caption{Optimized structures of the alanine tripeptide.
Different geometries correspond to the minima on the potential energy surface
(see contour plot in figure \ref{map_ala_x3}).}
\label{ala_x3_minima}
\end{figure}

In figure \ref{ala_x3_minima} we show the optimized structures of the alanine
tripeptide.  Different geometries correspond to the minima on the potential energy surface
(see contour plot in figure \ref{map_ala_x3}).

\begin{figure}[h]
\includegraphics[scale=0.7,clip]{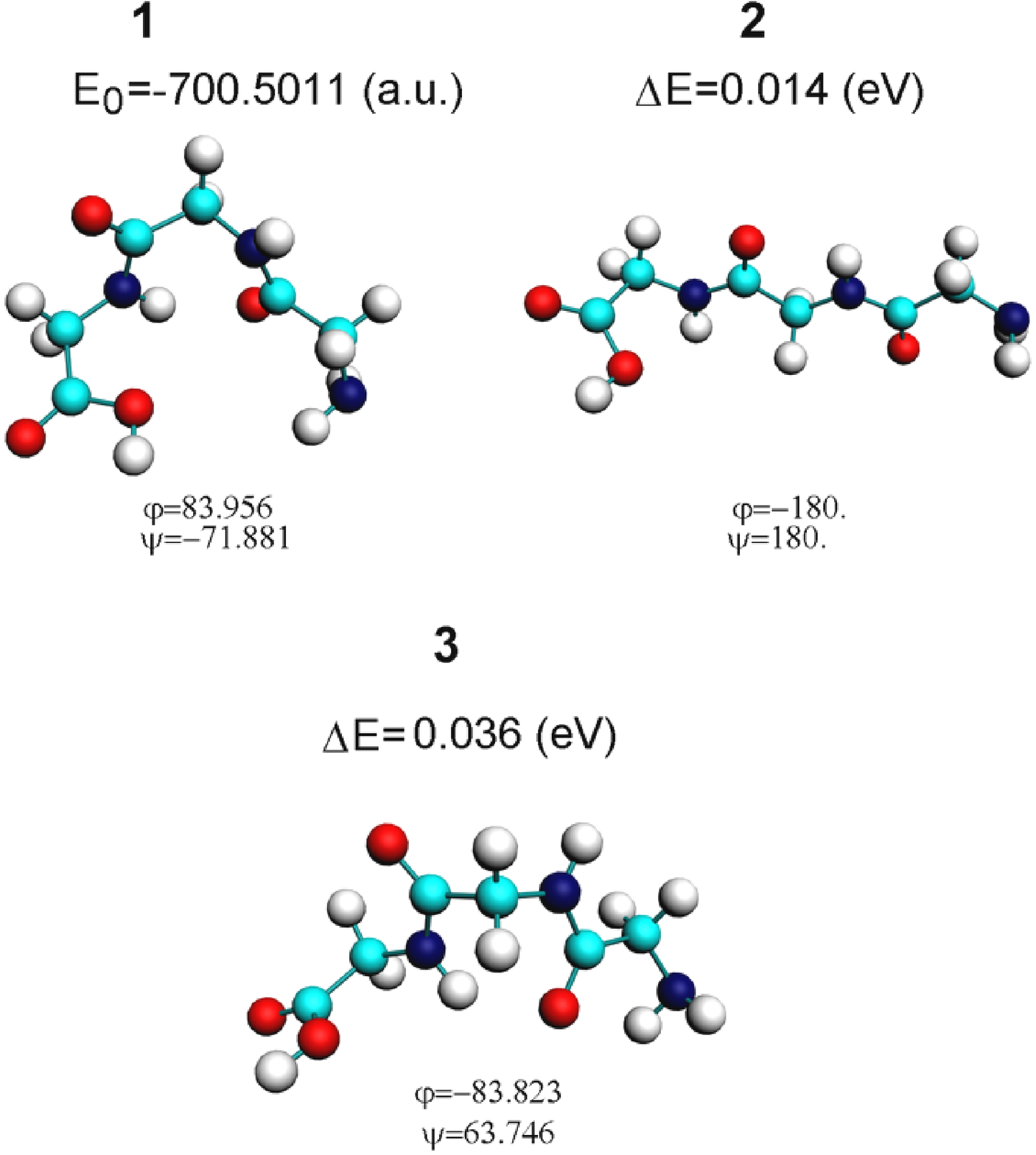}
\caption{Optimized structures of the glycine tripeptide.
Different geometries correspond to the minima on the potential energy surface
(see contour plot in figure \ref{map_gly_x3}).}
\label{gly_x3_minima}
\end{figure}

In figure \ref{gly_x3_minima} we show the optimized structures of the glycine
tripeptide.  Different geometries correspond to the minima on the potential energy surface
(see contour plot in figure \ref{map_gly_x3}).

\begin{figure}[h]
\includegraphics[scale=0.7,clip]{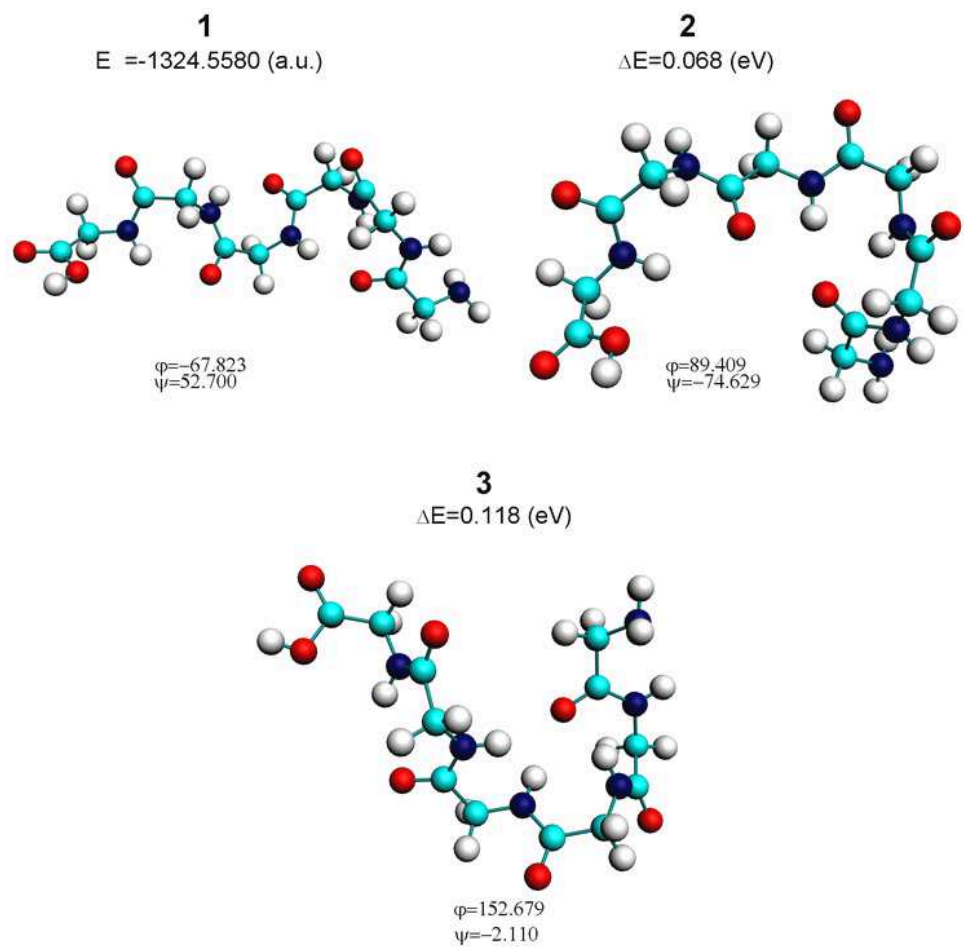}
\caption{Optimized structures of the glycine
hexapeptide in helix conformation.  Different geometries correspond to the minima on the potential
energy surface (see contour plot in figure \ref{map_gly_x6_helix})}
\label{gly_x6_minima_helix}
\end{figure}

In figure \ref{gly_x6_minima_helix} we show the optimized structures of the glycine
hexapeptide in helix conformation.  Different geometries correspond to the minima on the potential
energy surface (see contour plot in figure \ref{map_gly_x6_helix}).

\begin{figure}[h]
\includegraphics[scale=0.7,clip]{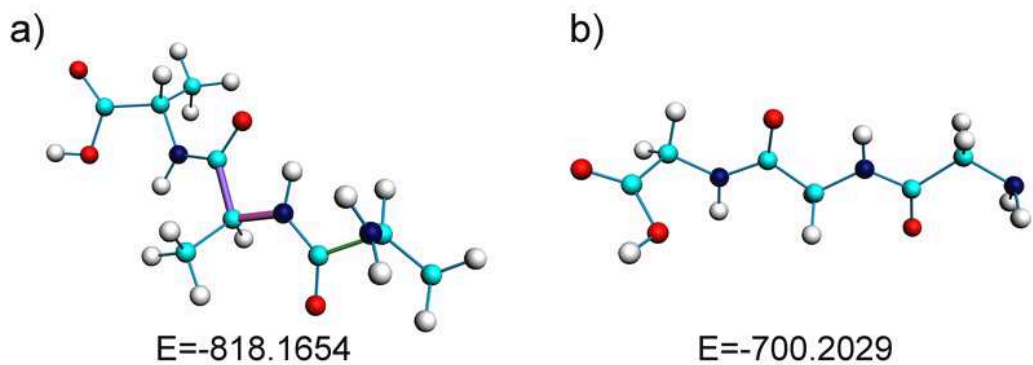}
\caption{Optimized geometries of singly charged alanine (a) and glycine (b) tripeptides
calculated with the B3LYP/6-31G(2d,p) method.
Energies below each image are given in a.u.}
\label{geom_charged}
\end{figure}

The geometries of singly charged alanine and glycine dipeptides are shown in figure
\ref{geom_charged}.

\begin{figure}[h]
\includegraphics[scale=0.8,clip]{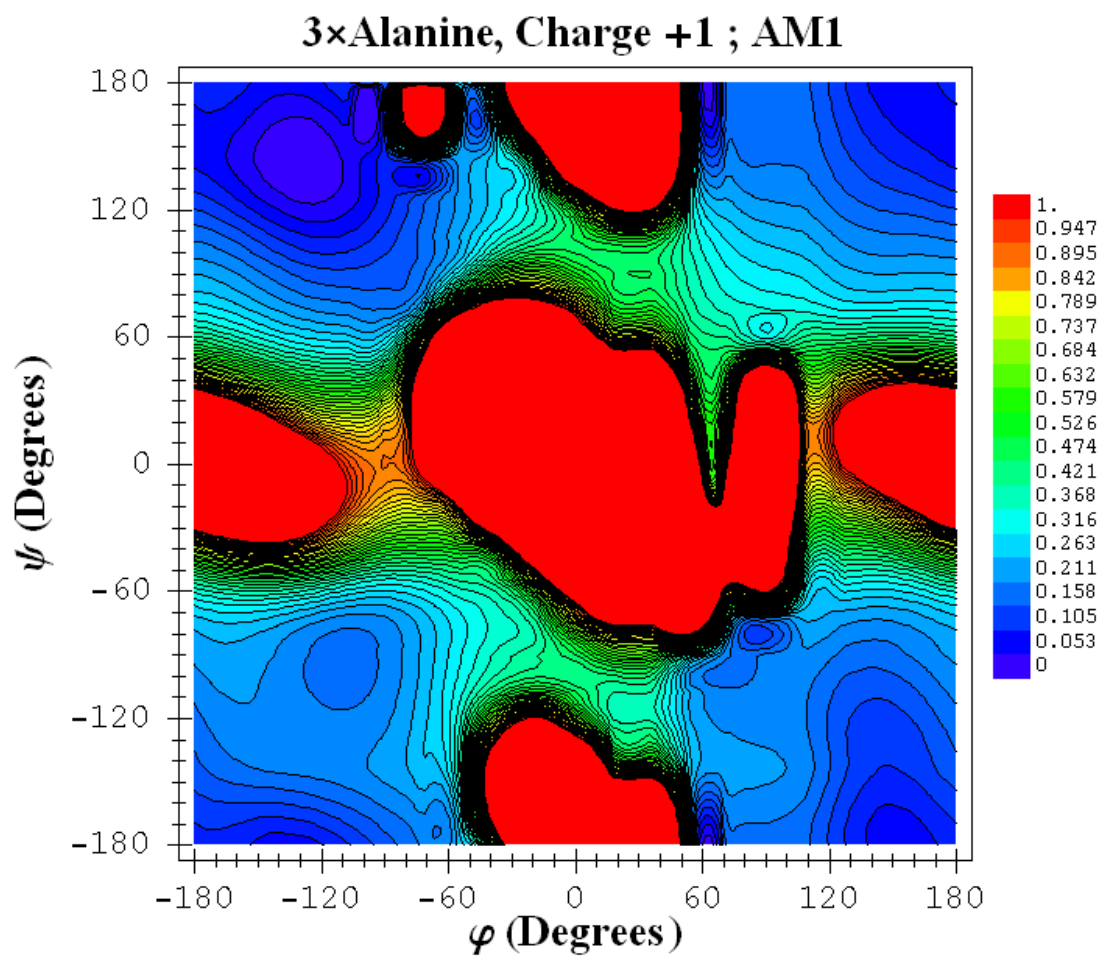}
\caption{Potential energy surface for the singly charged alanine tripeptide
calculated with the AM1 method.
Energies are given in eV.}
\label{ala_x3_+1_am1}
\end{figure}

\begin{figure}[h]
\includegraphics[scale=0.8,clip]{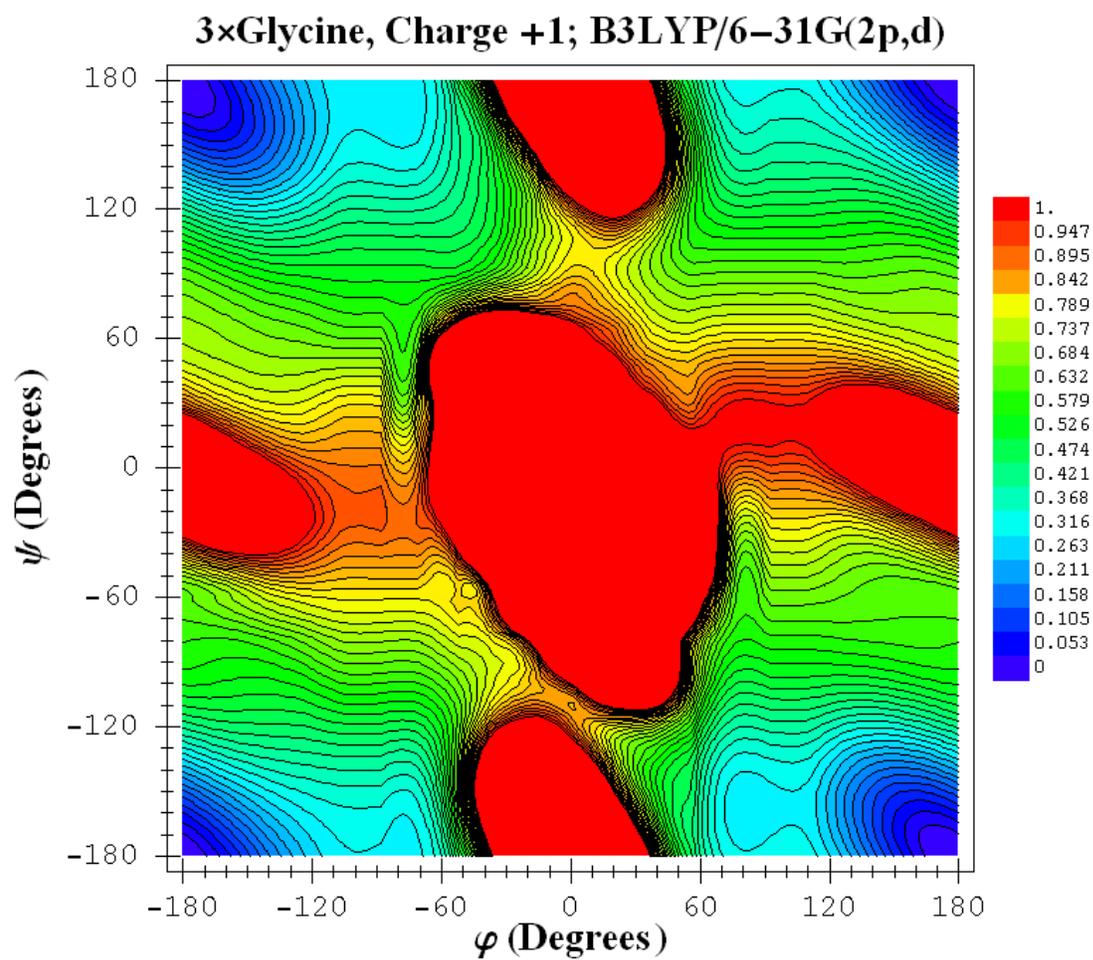}
\caption{Potential energy surface for the singly charged glycine tripeptide
calculated with the B3LYP/6-31G(2d,p) method.
Energies are given in eV.}
\label{gly_x3_+1}
\end{figure}

The ionization of the system changes dramatically its potential energy surface and the
secondary structure as it is seen from the contour plots presented in figures
\ref{ala_x3_+1_am1} and \ref{gly_x3_+1}.

\begin{figure}[h]
\includegraphics[scale=0.8,clip]{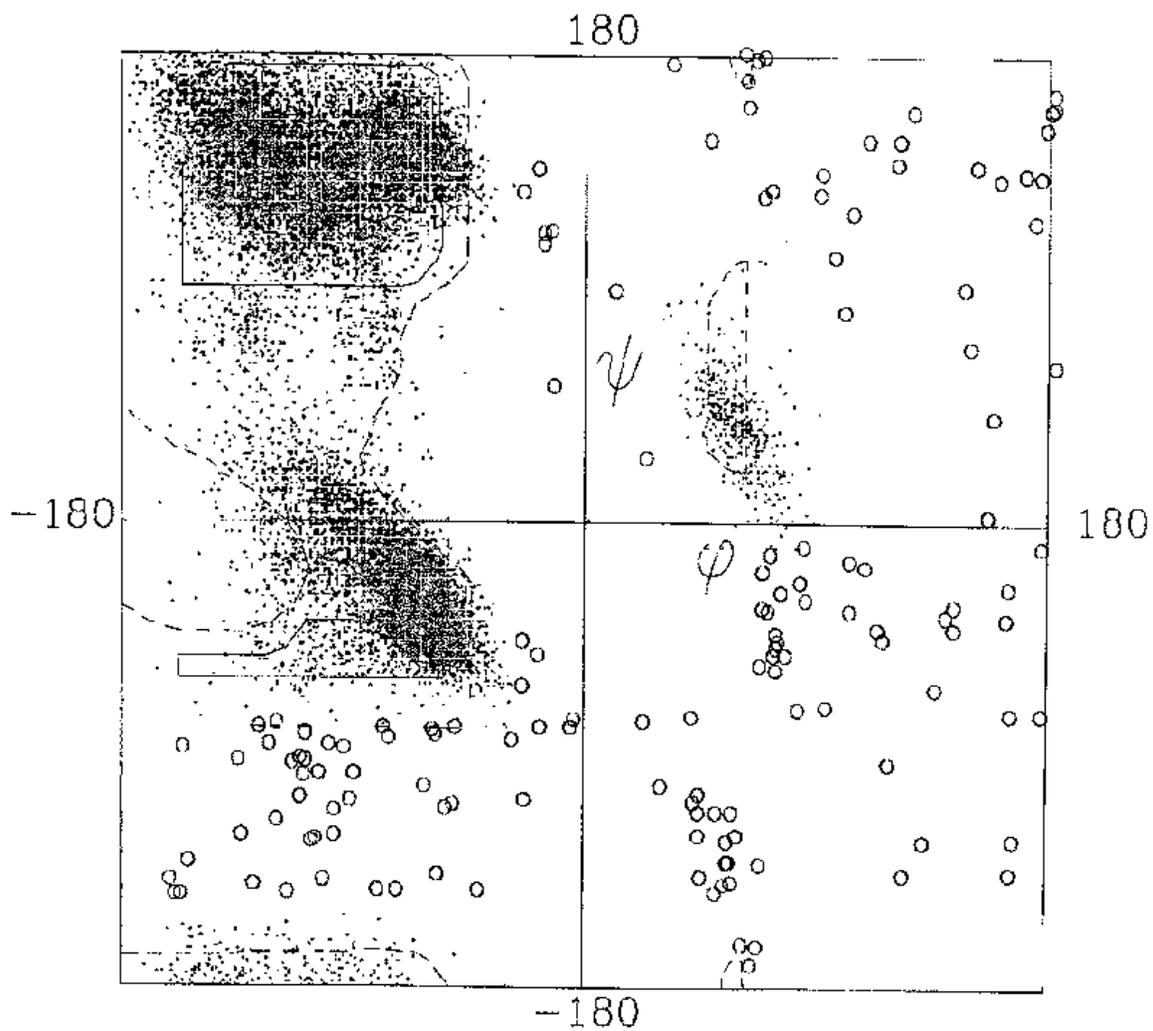}
\caption{Distribution of observed dihedral angles
$\varphi$, $\psi$ of non-Glycine residues in
protein structures selected from the Brookhaven Protein Data Bank
\cite{Gunasekaran96,Protbase}.}
\label{ramachandranEXP}
\end{figure}

In figure \ref{ramachandranEXP} we show the distribution of observed dihedral angles
$\varphi$, $\psi$ of non-Glycine residues in
protein structures selected from the Brookhaven Protein Data Bank
\cite{Gunasekaran96,Protbase}.
The circles show  conformations corresponding to the forbidden regions.

\newpage

\newpage

\newpage

\newpage

\newpage

\newpage

\newpage

\newpage

\newpage

\newpage

\newpage

\newpage

\newpage

\newpage

\newpage

\newpage

\newpage

\newpage

\newpage

\end{document}